\documentstyle[12pt]{article}

\begin{document}

\title{Current-Carrying Zero Mode for the Nielsen-Olesen String }
\author{V.B. Svetovoy \thanks{%
e-mail: vs@ics.ac.ru; tel. (0852) 11 22 81; fax (0852) 11 65 52}  \\
{\small Department of Physics, Yaroslavl State University,} \\ {\small %
Sovetskaya 14, Yaroslavl 150000, Russia}}
\date{}
\maketitle

\begin{abstract}
Zero modes of strings in the abelian Higgs model are analyzed. In spite of
the fact that the gauge symmetry is not broken in the string center, the
corresponding zero mode is shown to exist and to see it one has to analyze
carefully the dependence on transverse coordinates for the excitations. The
analysis of this kind is also important for the Witten model of
superconducting string. Unusual properties of the zero modes connected with
the broken gauge symmetry in the string background are investigated. One of
the modes carries the current quite similar to that in the Witten model and
gives back reaction to the string profile. It is claimed that the current in
the string improves stability of the electroweak string.
\end{abstract}


 \vspace{0.5cm}

Cosmic strings are one dimensional topological defects which can appear in
some spontaneously broken gauge theories during the phase transition \cite
{K76}. The strings predicted by many Grand Unified Theories are important
for cosmological applications \cite{HK}. They have been proposed as seeds
for large scale structure formation \cite{V85} and as a means to reproduce
temperature fluctuations in the cosmic microwave background \cite{AS90}.

Any string solution inevitably breaks some internal and space-time
symmetries generating massless excitations of the string ("zero modes")
connected with the broken transformations. In some cases these excitations
are very important for the string dynamics. A well known example is the
string superconductivity caused by the bosonic zero mode originated from the
broken electromagnetic gauge symmetry inside the string core \cite{W85}. In
the original Witten model the $\widetilde{U}(1)\times U(1)$ local symmetry
is broken by the string solution, where the first factor gives rise to the
Nielsen-Olesen vortex \cite{NO} and the second one (electromagnetic) becomes
involved via interaction of the $\widetilde{U}(1)$ and $U(1)$ charged
scalars. If the broken $U(1)$ symmetry generates a zero mode then why we
should not expect a similar mode for the broken $\widetilde{U}(1)$ symmetry?
This question was discussed in ref.\cite{KS92} where a whole spectrum of the
zero modes corresponding to the broken gauge symmetry and space-time
rotations was described for the abelian Higgs model. However, possible
physical manifestations of these modes have not been considered so far.

In this paper the physical role of the vortex zero modes in the abelian
Higgs model will be analyzed. It will be shown that if the field equations
are true for arbitrary distances, then these modes do not affect the
physics. However, if there is a short-distance cutoff ( the Plank scale $%
M_P^{-1}$ is a natural candidate ), one of them survives providing the
hypercharge current along the string. We will see that this mode gives back
reaction to the string forming fields and improves stability of the
electroweak string.

To start with, one considers more complicated but familiar example that is
the $\widetilde{U}(1)\times U(1)$ superconducting string \cite{W85}.
Concerning the last $U(1)$ factor in this model there is a charged scalar
field $\sigma $ which is condensed in the string core $\sigma =\overline{%
\sigma }(r)$ and an electromagnetic field $A_\mu $ vanishing for the string
solution $A_\mu =\overline{A}_\mu =0$. Given a condensate, there is, in
fact, a family of solutions for if $\sigma =\overline{\sigma }(r)$ is a
solution of the field equations so is $\sigma =\overline{\sigma }(r)\exp
(i\alpha )$. The phase is the Namby-Goldstone boson of the electromagnetic $%
U(1)$ symmetry, broken in the core of the string \cite{HK}. For the string
along the $z$-axis the low-energy excitations have been written as \cite{W85}

\begin{equation}
\label{az0}\sigma =\exp \left( ie\varphi (t,z)\right) \overline{\sigma }(r),
\end{equation}

\noindent where $\varphi (t,z)$ is arbitrary slowly varying function, $e$ is
the electric charge of $\sigma $, and $r,\vartheta $ are the polar
coordinates in the plain orthogonal to the $z$-axis. $\varphi (t,z)$ would
be an exact zero mode if the $U(1)$ symmetry were not gauged, but for the
local symmetry such an ansatz introduces nonzero current components $j_0$
and $j_z$ breaking the equations of motion for $A_0$ and $A_z$. It was
assumed further that $A_{0,z}$ are the functions of $t$ and $z$ only
whenever $\overline{\sigma }\neq 0$ and the problem was reduced to a
2-dimensional action for $\varphi (t,z)$ and $A_{0,z}(t,z)$. This is
conventional logic introducing the current-carrying zero mode. It is obvious
that this description is not absolutely correct since $A_{0,z}$ must depend
on $r$ whenever $\overline{\sigma }(r)$ changes rapidly. Accurate analysis
taking into account this dependence reveals unexpected properties of this
zero mode.

Corrected ansatz \cite{ABC} includes $r,\vartheta $ dependent phase $\varphi
(t,z;r,\vartheta )$ and excitation of the electromagnetic field. In the
coordinate notations $x^i=(t,z)$ for $i=0,1$ and $y^\alpha =\left( r\cos
\vartheta ,r\sin \vartheta \right) $ for $\alpha =1,2$ it is

\begin{equation}
\label{az1}\sigma =\exp \left( ie\varphi (x,y)\right) \overline{\sigma }%
(r),\quad A_\mu =\delta _\mu ^\alpha \partial _\alpha \varphi (x,y),
\end{equation}

\noindent where the equations of motion for the function $\varphi (x,y)$ are

\begin{equation}
\label{zmeq}\partial _\alpha ^2\varphi -2e^2\overline{\sigma }^2(r)\varphi
=0,
\end{equation}

\begin{equation}
\label{weq}\partial _i\partial ^i\varphi =0.
\end{equation}

\noindent Gauge transforming by $\chi =-\varphi $ gives the alternative
formulation

\begin{equation}
\label{az2}\sigma =\overline{\sigma }(r),\quad A_\mu =-\delta _\mu
^i\partial _i\varphi (x,y).
\end{equation}

\noindent which shows that the zero modes are the excitations of the $i$%
-components (i.e. $t$ and $z$) of the gauge field.

Such a change in the description would be elaboration of a good qualitative
description if not one circumstance: the solutions of eq.(\ref{zmeq}) are
always singular at $r\rightarrow 0$ or $r\rightarrow \infty $. On this basis
it was concluded \cite{ABC} that the energy of the modes diverges and there
must be a cutoff parameter for the theory to make sense.

A strict reader, probably, will propose here to exclude such excitations at
all since they do not satisfy to the boundary conditions. However, they do.
This question has been specially analyzed \cite{KS92,KS94} and it was shown
that for some class of massless excitations against nontrivial classical
backgrounds the boundary conditions are relaxed allowing some singular
behavior. In contrast with \cite{ABC} the energy of these modes was found to
be finite. Usual excitations of the string are restricted at $r\rightarrow 0$
and vanish at $r\rightarrow \infty $. The modes under consideration extend
the function space by generalized functions which are somewhat similar to
the $\delta $-function: they allow singularities but their integral effect
(energy) is finite.

To be more precise, one considers the abelian Higgs model where all the
described effects also take place. The lagrangian of a system consisting of
the $U(1)$ gauge field $B_\mu $ and the complex scalar field $\Phi $ is

\begin{equation}
\label{L}{\cal L}=-\frac 14F_{\mu \nu }F^{\mu \nu }+(D_\mu \Phi )^{*}(D_\mu
\Phi )-\lambda \left( \left| \Phi \right| ^2-\eta ^2/2\right) ^2,
\end{equation}

\noindent where

$$
F_{\mu \nu }=\partial _\mu B_\nu -\partial _\nu B_\mu ,\quad D_\mu \Phi
=\left( \partial _\mu -igB_\mu \right) \Phi .
$$

\noindent One takes a Nielsen-Olesen string configuration \cite{NO} given by

\begin{equation}
\label{NOstr}B_i=0,\quad B_\alpha =\frac 1{r^2}\varepsilon _{\alpha \beta
}y_\beta B(r),{\cal \quad }\Phi =\Phi _0(r)e^{i\vartheta }.
\end{equation}

\noindent At the boundaries the functions $B(r)$ and $\Phi _0(r)$ behave as

$$
r\rightarrow 0:\qquad B(r)\rightarrow 0,\quad \Phi _0(r)\rightarrow 0;
$$

\begin{equation}
\label{bc}r\rightarrow \infty :\qquad B(r)\rightarrow \frac 1g,\quad \Phi
_0(r)\rightarrow \frac \eta {\sqrt{2}}.\quad
\end{equation}

The string properties are often treated on the basis of trapped false vacuum
properties. The consequence of this approach is the conclusion that in the
abelian Higgs model the gauge symmetry is unbroken in the string core and so
there is no corresponding zero mode. However, such an approach is not quite
correct. For example, on this way it would not be possible to predict the
translational zero modes. Failure of the ansatz (\ref{az0}) is also the
result of this approach. The string fields interpolate between false and
true vacuua and definitely break the gauge symmetry so as the translations
in $y_\alpha $ directions and the rotations between $x_i$ and $y_\alpha $
coordinates.

The translational zero modes are localized near the string core, regular and
well known. They are not very interesting for our analysis. On the contrary,
the modes connected with the broken gauge symmetry (gauge modes) and those
arising as a result of the broken rotations (rotational modes) appear as
very unusual subjects. A common method for these modes construction has been
developed in ref.\cite{KS94}. Let $V^\mu (x,y)$ and $\phi (x,y)$ be the
excitations of the string fields $B^\mu $ and $\Phi $. The gauge can be
chosen so that the rotational and gauge modes are contained only in $V^i$
\cite{KS94} (see also (\ref{az2})). Massless excitations $V^i(x,y)$ obey the
equation

\begin{equation}
\label{zmeq1}\partial _\alpha ^2V_i-2g^2\Phi _0^2(r)V_i=0
\end{equation}

\noindent which, roughly speaking, is the condition of masslessness\footnote{%
This equation has been deduced \cite{KS94} analyzing the increase of the
background fields under broken symmetry transformations.}. Gauge modes are
the excitations solving eq.(\ref{zmeq1}) and having the gradient form

\begin{equation}
\label{gm}V^i(x,y)=\partial ^i\varphi (x,y).
\end{equation}

\noindent The rotational modes also have a special form

\begin{equation}
\label{rm}V^i(x,y) = \delta g^{i\alpha}(x,y)B_{\alpha}(y),
\end{equation}

\noindent where $\delta g^{i\alpha}(x,y)$ is the metric variation under
infinitesimal transformations of the coordinates. $V^i(x,y)$ one can expand
in Fourier series

\begin{equation}
\label{ftr}V^i(x,y)=\sum_{m=0}^\infty V_{m\alpha }^i(x)e_\alpha
^{(m)}(\vartheta )f_m(r),
\end{equation}

\noindent where $e_\alpha ^{(m)}(\vartheta )=(\cos m\vartheta ,\sin
m\vartheta )$ and the functions $f_m(r)$ satisfy the equation

\begin{equation}
\label{fmeq}f_m^{\prime \prime }+\frac 1rf_m^{\prime }-\left( 2g^2\Phi
_0^2(r)+\frac{m^2}{r^2}\right) f_m=0.
\end{equation}

The solutions of this equation inevitably contain singularities at the
origin or at infinity. Because of this special attention has to be paid to
the boundary conditions for eq.(\ref{zmeq1}). Typically boundary conditions
for excitations come from the demand that the surface term in first
variation of the action disappears. Any excitation $V^i(x,y)$ (massive or
massless) is not restricted by this condition since the background
components $B^i$ vanish. The boundary conditions on $V^i$ appear then from
the demand of finiteness for the second variation of the action which is

\begin{equation}
\label{S2}S^{(2)}=\int d^2xd^2y\left[ -\frac 14\left( \partial
_iV_j-\partial _jV_i\right) ^2+\frac 12\left( \partial _\alpha V_i\right)
^2+g^2\left| \Phi \right| ^2V_i^2\right] .
\end{equation}

\noindent For massless rotational or massive modes one finds the restriction

\begin{equation}
\label{restr}\int d^2xd^2y\partial _\alpha \left( V_i\partial _\alpha
V^i\right) =0
\end{equation}

\noindent which bounds massive modes at the boundaries in usual way and
there is a nontrivial singular solution of eq.(\ref{restr}) for the
rotational zero mode \cite{KS92,KS94}. The rotational mode needs special
discussion that will be done elsewhere. Let us concentrate here on the gauge
modes (\ref{gm}). For them the first term in (\ref{S2}) disappears and the
rest ones give rise to the kinetic terms of the 2-dimensional fields $%
\varphi _{m \alpha}(x)$. In this case (\ref{restr}) can be realized
restricting the $x$-dependence in usual way, $\int d^2x\partial _i(\varphi
_{m\alpha }\partial ^i\varphi _{m\alpha })=0$, instead of bound the $y$%
-behavior of the modes.

Substituting the expansion (\ref{ftr}) into the action (\ref{S2}) and
integrating it over the $y_\alpha $ coordinates one finds a 2-dimensional
lagrangian for the gauge modes

\begin{equation}
\label{L2}{\cal L}^{(2)}=\sum\limits_{m=0}^\infty \frac{Z_m}2\left( \partial
^i\varphi _{\alpha m}(x)\right) ^2,
\end{equation}

\noindent where $V_{m\alpha }^i(x)=\partial^i \varphi_{\alpha m}(x)$ and the
coefficients $Z_m$ are

\begin{equation}
\label{nc}Z_m=\pi (1+\delta _{0m})\int dr\frac d{dr}\left( rf_mf_m^{\prime
}\right) .
\end{equation}

\noindent This is what was expected for massless excitations. Although the
integral in (\ref{nc}) diverges, the action is not infinite because one has
to normalize $\varphi _{\alpha m}(x)$. The fields $\varphi _{\alpha m}(x)$
will be defined as 2-dimensional fields if the coefficients $Z_m=1$. This is
usual normalization condition for eigenfunctions which has to be imposed on
any kind of excitations. Since the functions $f_m(r)$ are singular at $%
r\rightarrow 0$, the normalization condition has to be defined with the help
of a cutoff parameter $r_0\rightarrow 0$:

\begin{equation}
\label{nc1}\pi (1+\delta _{0m})\int\limits_{r_0}^\infty dr\frac d{dr}\left(
rf_mf_m^{\prime }\right) =1.
\end{equation}

\noindent In contrast with the case of regular eigenfunctions (\ref{nc1})
can be considered as a boundary condition which together with eq.(\ref{fmeq}%
) completely defines the functions $f_{m}(r)$. On the other hand the
condition (\ref{nc1}) introduces generalized functions. Really, the set of
solutions of eq.(\ref{fmeq}) normalized according (\ref{nc1}) for $r_0\leq
r\ll R$ has the form

$$
f_0(r)=\frac{\ln (R/r)}{\sqrt{2\pi \ln (R/r_0)}},
$$

\begin{equation}
\label{gf}f_m(r)=\frac 1{\sqrt{\pi m}}\left( \frac{r_0}r\right) ^m,\qquad
m>0,
\end{equation}

\noindent where $R\sim g\eta $ is the string radius. Similar to the $\delta $%
-function this functions vanish at any finite $r$ and take some value at $%
r=r_0\rightarrow 0$. The integral condition for the $\delta $-function is
replaced by the normalization condition (\ref{nc1}). So, our zero modes live
strictly in the string center and obviously carry finite energy. The
conclusion of ref.\cite{ABC} that the energy of similar modes diverges was
made because the normalization condition was not imposed and the energy was
defined up to an arbitrary factor.

The $m=0$ mode induces charge per unit length and current of $U(1)$ charges
along the string. The current density is

\begin{equation}
\label{cd}j_i=-2g^2\partial _i\varphi (x,y)\Phi _0^2(r).
\end{equation}

\noindent Integrating $j_1$ over the string section with the help of (\ref
{ftr}) and (\ref{fmeq}) one finds for the current along the string

\begin{equation}
\label{cur}J=\int j_1d^2y=\partial _1\varphi _0(x)\sqrt{\frac{2\pi }{\ln
(R/r_0)}},
\end{equation}

\noindent where the function $\varphi _0(x)$ so as any $\varphi _{m\alpha
}(x)$ obeys the 2-dimensional wave equation. Note that the integral in (\ref
{cur}) is defined by the asymptotics of $f_0(r)$ but not the condensate
distribution $\Phi _0(r)$. The same result (\ref{cur}) will be found for the
electric current in the Witten model in spite of difference in the
condensates distribution. It differs from the expression found in \cite{W85}
by the numerical factor but the qualitative result does not change. The
reason for this difference, as was explained before, is taking into account $%
r$-dependence for the zero mode.

In the limit $r_0\rightarrow 0$ the current $J$ goes to zero. However, in
physical reality we cannot believe in our equations for arbitrary small
distances. A natural candidate for the cutoff parameter $r_0$ is the Plank
scale $M_P^{-1}$. With such a cutoff the current remains practically
unsuppressed even if the string was formed at the electroweak transition
when $R\sim M_W^{-1}$. The gauge modes are concentrated near the string
center because of their singular nature. Indeed, all the $m>0$ modes cannot
influence low energy physics since they vanish rapidly at a distance from
the center large in comparison with $M_P^{-1}$. In contrast with that the $%
m=0$ mode decreases logarithmically and survives at large distances.
Existence of the current itself is the direct result of the short distance
behavior.

One suppose $\varphi _0(t,z)$ varies linearly with time and position

\begin{equation}
\label{gszm}\varphi _0(t,z)=b_0t-b_1z.
\end{equation}

\noindent In this case the string is described by the time and position
independent fields

\begin{equation}
\label{az3}B_i=b_if_0(r),\qquad B_\alpha =\frac 1{r^2}\varepsilon _{\alpha
\beta }y_\beta B(r),\qquad \Phi =\Phi _0(r)e^{i\vartheta },
\end{equation}

\noindent and can be considered as a ground state with charge and current.
The zero mode gives back reaction to the string profile because the equation
for $\Phi _0(r)$ includes an additional term $g^2(b_0^2-b_1^2)f_0^2\Phi _0$.

The Nielsen-Olesen vortex solution can be embedded into the $SU(2)_L\times
U(1)_Y$ electroweak theory \cite{V92} and it is known as electroweak string.
However, there are no topological arguments for stability of this string and
it happens so that dynamical stability is reached outside of the physical
region of parameters \cite{JPV93}. Perturbation spectrum for the electroweak
string in presence of charge and current is modified. The problem will be
analyzed elsewhere but to see the tendency one makes a few comments on the
case $\sin ^2\theta _W=1$ (semilocal string \cite{VA91}). The string profile
is described by eq.(\ref{az3}), where $\Phi $ is the down component of the
Higgs doublet and one has to replace $B_\mu \rightarrow Z_\mu $ and $%
g\rightarrow q$. One can show that in our case the stability analysis can be
reduced to an analysis of the excitations in the upper component of the
Higgs doublet $\phi $. Since the background fields do not depend on $t$ and $%
z$, $\phi $ can be sought in the form

\begin{equation}
\label{az4}\phi (x,y)=e^{i\omega t}e^{-ikz}\phi (r,\vartheta ).
\end{equation}

\noindent Expanding $\phi (r,\vartheta )$ in angular momentum states $\phi
_m(r)e^{im\vartheta }$ one comes to the eigenvalue problem for $\omega $

$$
\phi _m^{\prime \prime }+\frac 1r\phi _m^{\prime }+\left[ \left( \omega
+qb_0f_0\right) ^2-\left( k+qb_1f_0\right) ^2\right] \phi _m-
$$

\begin{equation}
\label{eign}\left[ \frac{(m-qB)^2}{r^2}+2\lambda \left( \left| \Phi \right|
^2-\eta ^2/2\right) \right] \phi _m=0
\end{equation}

\noindent One can immediately conclude that the current ($b_1\neq 0$)
improves string stability but the charge on the string ($b_0\neq 0$) acts in
the opposite direction. Therefore, it is expected that stability of the
current-carrying electroweak strings will be improved.

In conclusion, it was shown that the current-carrying zero mode exists in
the abelian Higgs model with the Nielsen-Olesen string solution. The
argument that the gauge symmetry is unbroken inside of the string core does
not work because the zero mode is connected rather with the transition
between vacuua inside and outside of the string. The mode has logarithmic
singularity in the string center, but it is normalizable in the sense of
generalized functions. The current along the string disappears as $1/\sqrt{%
\ln (R/r_0)}$ when the short distance cutoff $r_0$ goes to zero. The
physical cutoff parameter $r_0\sim M_P^{-1}$ leaves the current practically
unsuppressed. It was claimed that the current in the string improves
stability of the electroweak string.

\vspace{5mm}

\noindent {\bf Note added}

\vspace{3mm}

After the submission of this paper author received a preprint by Kibble {\it %
et al} \cite{KLY97} in which a nonabelian current-carrying string
configuration similar to (\ref{az3}) is considered. They investigated the
gauge zero mode which has logarithmic singularity at infinity and instead of
the normalization condition for the mode have connected behavior at large
distances with the current in the string.

\end{document}